\documentclass[12pt]{article}

\usepackage{bbm,amsmath}
\usepackage{amscd}
\usepackage{amssymb}
\usepackage{amsfonts}
\usepackage[mathscr]{euscript}
\usepackage{pstricks}
\usepackage{pst-node}
\usepackage{graphicx}

\usepackage{amssymb, theorem}
\usepackage[latin2]{inputenc}
\usepackage{graphicx}

\def\iS{\mathcal S}

\def\i1{\mathcal 1}

\def\1{\mathbbm{1}}

\def\>{\rangle}
\def\<{\langle}

\begin{document}

\title{\bf Three qubits in a symmetric environment: dissipatively generated asymptotic 
entanglement}

\author{Fabio Benatti$^{a,b}$, Adam Nagy$^c$\\
\small $^a$Dipartimento di Fisica, Universit\`a di Trieste,
Strada Costiera 11,\\
\small 34051 Trieste, Italy\\
\small $^b$Istituto Nazionale di Fisica Nucleare, Sezione di
Trieste, 34100 Trieste, Italy\\
\small $^c$Budapest University of Technology and Economics,
Muegyetem rkp.3-9, Hungary}

\date{\null}

\maketitle

\begin{abstract}
We study the asymptotic entanglement of three identical qubits under the action
of a Markovian open system dynamics that does not distinguish them.
We show that by adding a completely depolarized qubit to a special class of two qubit states, by letting them reach the asymptotic state and by finally eliminating the added qubit, 
can provide more entanglement than by direct immersion of the two qubits within the same environment.
\end{abstract}

\section{Introduction}

A quantum system $S$ interacting weakly with its environment $E$ can be treated as an open quantum system: standard techniques can be applied to obtain a master equation involving the degrees of freedom of $S$, only.
On a long time-scale determined by the weakness of the coupling to the environment, the reduced dynamics consists of a semigroup of trace-preserving completely positive
maps on the states of $S$. These maps are fully consistent transformations of the system $S$ states
incorporating the dissipative and noisy effects due to the environment $E$~\cite{AL,Pe}.

An irreversible reduced dynamics typically transforms pure states into mixed states, thus
spoiling fundamental quantum resources like entanglement~\cite{Bruss}.
However, the Markovian reduced dynamics resulting from suitably engineered environments
may entangle an initial separable state of a bipartite quantum system immersed
within it; such an entanglement can even persist asymptotically~\cite{braun,beige,jakob1,jakob2,BFP,BFrev,BLN,Isar,Kraus}.

The properties of the asymptotic states of quantum dynamical semigroups can be studied by looking at
the structure of the generator of the reduced dynamics~\cite{Fri1,Spohn,Fri2,FagnRebol,Dietz,HBW,HB}:
some concrete applications can be found in~\cite{BFrev,Kraus,BF1,BF2,BF3}.

Most of the results present in the literature concern pairs of open qubits; in the following, we will instead consider three qubits weakly interacting with a common environment such that the
resulting master equation affects all possible pairs of qubits in the same way.
In this case, the manifold of stationary states can partially be analytically characterized
as well as the asymptotics of a particular class of initial states.
We shall use the analytically obtained results to construct the following protocol:
given two qubits immersed in the just depicted environment, 1)
we add a third completely depolarized qubit, 2) leave the three of them irreversibly
evolve until they reach the stationary state, 3) trace away the ancilla.
As an indication of the surprises set in store by higher dimensional entanglement with respect to 
the two qubit case, we show that, for certain initial two-qubit states, the entanglement obtainable by such a procedure is larger than that achievable by simply letting them reach the stationary state within the bath.

\section{Three open qubit model}

In the following we shall study an open quantum system of three qubits
in weak interaction with their environment $E$.
We shall denote by $M$ the algebra of $8\times 8$ matrices $x\in M_8(\mathbb{C})$ and by
$\rho\in M$ the positive matrices of trace $1$ that describe the states of $S$ and
by $\iS(M)$ the convex set of all states.
For sake of simplicity, we shall sometimes write $\rho(x)$ for the expectation values ${\rm Tr}(\rho\,x)$;
further, by $x^{(1)},y^{(2)},z^{(3)}$, respectively $\rho_1^{(1)},\rho_2^{(2)},\rho_3^{(3)}$,
we will denote the local one-qubit observables
$x\otimes \mathbbm{1}\otimes\mathbbm{1}$, $\mathbbm{1}\otimes y\otimes\mathbbm{1}$, $\mathbbm{1}\otimes\mathbbm{1}\otimes z$, respectively the one-qubit states $\rho_1\otimes \mathbbm{1}\otimes\mathbbm{1}$, $\mathbbm{1}\otimes\rho_2\otimes\mathbbm{1}$ and $\mathbbm{1}\otimes\mathbbm{1}\otimes\rho_3$.

In the weak-coupling limit, the states of an open quantum system
evolve in time according to a master equation of the form $\partial_t\rho_t=L[\rho_t]$, where the generator $L$
incorporates the dissipative and noisy effects due to the environment; the solutions form a semigroup of
completely positive maps $\gamma_t=\exp(t\,L)$, $t\geq 0$~\cite{AL,Pe,BFrev}.
In the present case of three qubits, we shall concretely study the following time-evolution equation,
\begin{equation}
\label{ME0}
\partial_t\rho_t=-i\Big[\sum_{a=1}^3\frac{\omega_a}{2}\,\sigma^{(a)}_3\,,\,\rho_t\Big]+
\sum_{a,b=1\atop i,j=1}^3C^{(ab)}_{ij}
\Big(\sigma^{(a)}_i\,\rho_t\, \sigma^{(b)}_j-\frac{1}{2}\,\Big\{\sigma^{(b)}_j\sigma^{(a)}_i
\,,\,\rho_t\Big\}\Big)=L[\rho_t]\ ,
\end{equation}
where $\sigma^{(a)}_i$ is the $i$-th Pauli matrix of the $a$-th qubit, and
the coefficients $C^{(ab)}_{ij}$ form a a $9\times 9$ positive matrix, the so-called Kossakowski matrix:
\begin{equation}
\label{K0}
0\leq K=\begin{pmatrix}C^{(11)}&C^{(12)}&C^{(13)}\cr
(C^{(12)})^\dag&C^{(22)}&C^{(23)}\cr
(C^{(13)})^\dag&(C^{(23)})^\dag&C^{(33)}
\end{pmatrix}\ ,\quad
C^{(ab)}=[C^{(ab)}_{ij}]_{i,j=1}^3\ .
\end{equation}
To the semigroup of completely positive, trace-preserving maps
$\gamma_t=\exp(t\,L)$ on the states of $S$, there corresponds
the semigroup of completely positive, identity-preserving maps
$\hat{\gamma}_t:M\mapsto M$, on the matrices $x\in M$, generated by:
\begin{equation}
\label{ME1}
\partial_tx_t=i\Big[\sum_{a=1}^3\frac{\omega_a}{2}\,\sigma^{(a)}_3\,,\,x_t\Big]+
\sum_{a,b=1\atop i,j=1}^3 C^{(ab)}_{ij}\,
\Big(\sigma^{(b)}_j\,x_t\, \sigma^{(a)}_i\,-\frac{1}{2}\,\Big\{\sigma^{(b)}_j\sigma^{(a)}_i
\,,\,x_t\Big\}\Big)=\hat{L}[x_t]\ .
\end{equation}

The analytic solution of~(\ref{ME0}) can in general be addressed only numerically;
we shall instead consider a simpler class of master equations amenable to a partially analytic study.
Notice that, by taking the trace of the generator $L$ in~(\ref{ME0}) with respect to any single qubit,
one gets the generator of a master equation relative to the other two qubits; thus, by choosing
$\omega_1=\omega_2=\omega_3=\omega$ and
$$
C^{(11)}=C^{(22)}=C^{(33)}=A\ ;\
C^{(12)}=C^{(21)}=C^{(13)}=C^{(31)}=C^{(23)}=C^{(32)}=B\ ,
$$
one obtains a highly symmetric generator with Kossakowski matrix
\begin{equation}
\label{SKM}
K=\begin{pmatrix}
A&B&B\cr B^\dag &A&B\cr B^\dag&B^\dag&A
\end{pmatrix}\ ,
\end{equation}
such that any pair of qubits is affected in the same way by the presence of the environment.
As apparent from~(\ref{SKM}), the matrix $A$ governs the dissipative reduced dynamics of each one of the qubits, while
$B$ rules the dissipative statistical coupling of pairs of different qubits.
However, in the following, we shall further restrict the case to master equations where one and two-qubit terms
are the same and choose
\begin{equation}
\label{K1}
A=B=\begin{pmatrix}
a&ib&0\cr
-ib&a&0\cr
0&0&c
\end{pmatrix}\ ,\qquad c\geq 0\ ,\ 0\leq |b|< a\ .
\end{equation}
Then, the Kossakowski matrix~(\ref{K0}) reads
$K=3\,A\otimes P$, where $P$ projects onto the vector $(1,1,1)/\sqrt{3}$ and $A\geq 0$.
The resulting master equation for the states (Schr\"odinger time-evolution) and its dual for the system operators (Heisenberg time-evolution) can conveniently be recast as
\begin{eqnarray}
\label{ME2}
\partial_t\rho_t&=&-i\frac{\omega}{2}\,\Big[S_3\,,\,\rho_t\Big]\,+\,
\sum_{i,j=1}^3 A_{ij}\,\Big(S_i\,\rho_t\,S_j\,
-\frac{1}{2}\,\Big\{S_j\,S_i\,,\,\rho_t\Big\}\Big)=L[\rho_t]\\
\label{ME3}
\partial_t x_t&=&i\frac{\omega}{2}\,\Big[S_3\,,\,x_t\Big]\,+\,
\sum_{i,j=1}^3 A_{ij}\,\Big(S_j\,x_t\,S_i\,
-\frac{1}{2}\,\Big\{S_j\,S_i\,,\,x_t\Big\}\Big)=\hat{L}[x_t]\ ,
\end{eqnarray}
in terms of global spin operators
\begin{equation}
\label{superspin}
S_{1,2,3}=\sum_{a=1}^3\sigma^{(a)}_{1,2,3}=\sigma^{(1)}_{1,2,3}\,+\,\sigma^{(2)}_{1,2,3}\,+\,\sigma^{(3)}_{1,2,3}\ .
\end{equation}
\medskip

\noindent
\textbf{Remark 1.}\quad
For three qubits ($n=3$), a master equation of the form
\begin{eqnarray*}
\partial_t\rho_t&=&-i\frac{\omega}{2}\,\Big[S_3\,,\,\rho_t\Big]+
\sum_{a=1}^n\sum_{i,j=1}^3 A_{ij}
\Big(\sigma^{(a)}_i\,\rho_t\, \sigma^{(a)}_j-\frac{1}{2}\,\Big\{\sigma^{(a)}_j\sigma^{(a)}_i
\,,\,\rho_t\Big\}\Big)\\
&+&
\sum_{a\neq b=1}^n\sum_{i,j=1}^3 B_{ij}
\Big(\sigma^{(a)}_i\,\rho_t\, \sigma^{(b)}_j-\frac{1}{2}\,\Big\{\sigma^{(b)}_j\sigma^{(a)}_i
\,,\,\rho_t\Big\}\Big)\ ,
\end{eqnarray*}
may have direct experimental implication
in certain realizations of the driven cavity array proposed in~\cite{Angelak}.
For $n=2$, the above equation have been derived in a physical scenario where the qubits are at a distance from each
and immersed in a scalar Bose field in thermal equilibrium~\cite{BF1}, while
a master equation of the form
$$
\partial_t\rho_t=-i\frac{\omega}{2}\,\Big[S_3\,,\,\rho_t\Big]+
\sum_{a,b=1}^3\sum_{i,j=1}^3\,A_{ij}
\Big(\sigma^{(a)}_i\,\rho_t\, \sigma^{(b)}_j-\frac{1}{2}\,\Big\{\sigma^{(b)}_j\sigma^{(a)}_i
\,,\,\rho_t\Big\}\Big)\ ,
$$
corresponds to two qubit immersed in an environment described by a thermal, scalar Bose field
when the spatial distance among the qubits is negligible~\cite{BFrev}.

\section{Asymptotic States}

We shall start by briefly reviewing some available results about the stationary states of quantum dynamical semigroups~\cite{Fri1,Spohn,Fri2} and about the tendency to equilibrium of open quantum systems.

Let $\iS_\gamma=\{\rho\in\iS(M)\,:\, \gamma_t[\rho]=\rho\ \forall t\geq0\}$ denote
the set of stationary states of a semigroup of trace-preserving, completely positive
maps $\gamma_t=\exp(tL)$, generated by the master equation $\partial_t\rho_t=L[\rho_t]$, and by
$M_\gamma=\left\{x\in M\,:\, \hat{\gamma}_t[x]=x\ \forall t\geq 0\right\}$
the set of operators ($n\times n$ matrices) invariant under the identity-preserving maps
$\hat{\gamma}_t=\exp(t\hat{L})$ generated by the dual time-evolution equation
$\partial_t x_t=\hat{L}[x_t]$.
It is convenient to cast the latter equation in diagonal form:
\begin{equation}
\label{KrausME}
\partial_t x_t=i\,\Big[H\,,\,x_t\Big]\,+\,
\sum_{i=1}^3\Big( V^\dag_i\,x_t\,V_i\,
-\frac{1}{2}\,\Big\{V^\dag_i\,V_i\,,\,x_t\Big\}\Big)\ .
\end{equation}
From~\cite{Fri2} one knows that, if a full-rank stationary state $\rho_\infty$ exists, then
\begin{enumerate}
\item
the subset of constant matrices $M_\gamma=\{x\in M:\ \hat{\gamma}_t[x]=x\}$ is a $*$-subalgebra of $M$, that
is $\hat{\gamma}_t[x^\dag]=x^\dag$ and also $\hat{\gamma}_t[x^\dag x]=x^\dag x$ for all $t\geq 0$.
\item
the time-average
\begin{equation}
\label{condexp}
\hat{\mathbb{E}}[x]=\lim_{T\to+\infty}\frac{1}{T}\,\int_0^T{\rm d}t\,\hat{\gamma}_t[x]
\end{equation}
defines a \textit{conditional expectation} from $M$ onto $M_\gamma$, that is a completely positive unital map
such that
\begin{equation}
\label{expcondprop}
\hat{\mathbb{E}}[1]=1\ ,\qquad
\hat{\mathbb{E}}[y_1\,x\,y_2]=y_1\,\hat{\mathbb{E}}[x]\,y_2\qquad\forall\,y_{1,2}\in M_\gamma\ ,\quad\forall\, x\in M\ .
\end{equation}
\end{enumerate}

The conditional expectation $\hat{\mathbb{E}}:M\mapsto M_\gamma$ has a dual map defined by
\begin{equation}
\label{dualcond}
{\rm Tr}\Big(\rho\,\hat{\mathbb{E}}[x]\Big)={\rm Tr}\Big(\mathbb{E}[\rho]\,x\Big)\ ,\qquad\forall\, \rho\in\iS(M)\ ,\ x\in M\ .
\end{equation}
This is a completely positive, trace-preserving linear map on the state-space $\iS(M)$ such that
$\mathbb{E}[\rho]$ is a stationary state and $\mathbb{E}[\rho]=\rho$ if $\rho$ is a stationary state.

We are interested in establishing whether, given any initial state $\rho$, it goes into an asymptotic
state $\rho_\infty$ according to
\begin{equation}
\label{limitstate}
\rho_\infty=\lim_{t\to+\infty}\gamma_t[\rho]=\mathbb{E}[\rho]\ .
\end{equation}
A sufficient condition can be obtained as follows: consider the subset $D_\gamma\subseteq M$ of $x\in M$ such that
$$
\hat{L}[x^\dag\,x]\,-\,\hat{L}[x^\dag]\,x\,-\,x^\dag\,\hat{L}[x]=0\ .
$$
From~(\ref{KrausME}) one immediately derives that $M_\gamma\subseteq D_\gamma$ and also that $x\in D_\gamma$ if and only if
$$
\sum_{i=1}^{3}([x\,,\,V_i])^\dag[x\,,\,V_i]=0\Longleftrightarrow
[x\,,\,V_i]=0\qquad\forall\, V_i\ .
$$
Thus, the subset $D_\gamma$ consists of $x\in M$ which commute with all operators $V_i$, namely
$D_\gamma=\{V_i\}'$ where $\{V_i\}'$ denotes the so-called commutant of the set $\{V_i\}$.
The commutant is a subalgebra which need not coincide with the time-invariant $*$-subalgebra $M_\gamma$.
This is however the case if the operators~\cite{Fri2} commuting with all $V_i$ also commute with their adjoints and with the Hamiltonian $H$. Indeed, if
$\{V_i\}'=\{V_i,V_i^\dag,H\}'$, then
$M_\gamma\subseteq D_\gamma\subseteq M_\gamma$ as~(\ref{KrausME}) yields
$\{V_i,V_i^\dag,H\}'\subseteq M_\gamma$.

Moreover, the equality $\{V_i\}'=\{V_i,V_i^\dag,H\}'$ is also sufficient~\cite{Fri2} to guarantee that
$\displaystyle
\lim_{t\to+\infty}\hat{\gamma}_t[x]=\hat{\mathbb{E}}[x]$, for all $x\in M$,
whence~(\ref{limitstate}) follows by duality.

In general, that is for any number of qubits, the time-evolution equation~(\ref{ME3}) can be written as in~(\ref{KrausME}) by diagonalizing the $2\times 2$ matrix in the upper left corner of $A$ in~(\ref{K1});
concretely, in terms of the spin operators $S_i$ in~(\ref{superspin}),
\begin{equation}
\label{Krausop}
V^\dag_{1,2}=\sqrt{2(a\mp b)}\,\frac{S_1\mp i\,S_2}{2}\ ,\qquad V_3=\sqrt{c}\, S_3\ .
\end{equation}
If $|b|<a$ and $c>0$, $\{V_i\}'=\{V_i,V_i^\dag,H\}'=\{S_i\}'$ so that, according to the above discussion, it follows that $M_\gamma=\{S_i\}'$. Therefore, in the concrete cases
we are considering, the time-invariant operators coincide with those commuting with all global spin operators $S_i$ in~(\ref{superspin}).
In order to establish the asymptotic convergence to stationary states as in~(\ref{limitstate}), we need seek full rank stationary states: we shall do this in the following for $1$, $2$ and $3$ qubits.

\subsection{One qubit}

For the case of one qubit, a full rank stationary state of the master equation
\begin{equation}
\label{ME5}
\partial_t\rho_t=-i\frac{\omega}{2}\,\Big[\sigma_3\,,\,\rho_t\Big]\,+\,
\sum_{i,j=1}^3 A_{ij}\,\Big(\sigma_i\,\rho_t\,\sigma_j\,
-\frac{1}{2}\,\Big\{\sigma_j\,\sigma_i\,,\,\rho_t\Big\}\Big)\ ,
\end{equation}
with $A=[A_{ij}]$ as in~(\ref{K1}), can be found by considering the corresponding time-evolution equation
of the Bloch vector $\vec{r}_t$ in
$\displaystyle\rho_t=\frac{1}{2}(1+\vec{r}_t\cdot\vec{\sigma})$; that is, $\dot{\vec{r}}_t=-2(\mathcal{L}\vec{r}_t-\vec{z})$,
where
\begin{equation}
\label{bloch}
\mathcal{L}=\begin{pmatrix}
a+c&-\omega/2&0\cr
\omega/2&a+c&0\cr
0&0&2a\end{pmatrix}\ , \quad\vec{z}=\,\begin{pmatrix}0\cr0\cr 2b\end{pmatrix}\ .
\end{equation}
Setting $\dot{\vec{r}}_t=0$, one finds $\vec{r}_\infty=\mathcal{L}^{-1}\vec{z}=(0,0,b/a)$
and a unique full-rank ($|b|<a$) stationary state
\begin{equation}
\label{blochstat}
\rho^*_\infty=\frac{1}{2}\Big(1+r_\infty\,\sigma_3\Big)\ ,\quad r_\infty=\frac{b}{a}\ .
\end{equation}

\noindent
\textbf{Remark 2.}\quad
Consider two qubits ($a=1,2$ in~(\ref{superspin})); one explicitly verifies (see also~\cite{BFrev}) that
$\rho^{\otimes 2}_\infty=\rho^*_\infty\otimes\rho^*_\infty$
is a full-rank stationary state for~(\ref{ME2}): $L[\rho^{\otimes 2}_\infty]=0$.

The generator of~(\ref{ME2}) can be extended to the case of $n$ qubits by extending to $n$ the 
summation index of single qubit Pauli matrices in~(\ref{superspin}); further, it can conveniently be recast as
$L=\sum_{a,b=1}^n\,L_{ab}$ where the sum is over generators~(\ref{ME2}) involving only the $a$th and $b$th qubit.
Let $\displaystyle \rho_\infty^{\otimes n}=\underbrace{\rho_\infty^*\otimes\rho_\infty^*\cdots\otimes\rho_\infty^*}_{n\, times}$; then,
$$
L^{(12)}[\rho_\infty^{\otimes n}]:=\Big(L_{11}+L_{22}+L_{12}+L_{21}\Big)[\rho^{\otimes n}_\infty]=
L^{12}[\rho^{\otimes 2}_\infty]\otimes\rho_\infty^{\otimes (n-2)}=0\ ,
$$
where $L^{12}$ is the generator in~(\ref{ME2}) for two qubits and
$\rho_\infty^{\otimes 2}$ is a two qubit stationary state.
This result clearly holds for all pairs $(ab)$, that is
$L^{(ab)}[\rho_\infty^{\otimes n}]=0$, whence
$L[\rho_\infty^{\otimes n}]=0$ and
$\rho^{\otimes n}_\infty$ is an $n$-qubit full-rank stationary state.

\subsection{Two qubits}

As previously observed, $\{V_i\}'=\{V_i,V^\dag_i,H\}'=\{S_i\}'$ independently of the number of qubits.
In order to find the commutant $\{S_i\}'$ for the case of two qubits, we use the Pauli matrices and write
$$
M\ni x=\lambda\,\mathbbm{1}\,+\,\sum_{i=1}^3\sum_{a=1}^2\lambda^{(a)}_i\,\sigma^{(a)}_i\,+\,
\sum_{i,j=1}^3\lambda_{ij}\,\sigma_i\otimes\sigma_j\ .
$$
Then, by imposing that $[x\,,\,S_p]=0$ for all $p=1,2,3$, $\{V_i\}'$ amounts to being
the linear span of the identity matrix $\mathbbm{1}$ and of the symmetric sum
$T=\sum_{i=1}^3\sigma_i\otimes\sigma_i$.
It follows that $M_\gamma=\{S_i\}'$ is a commutative algebra; it coincides with its center,
$M_\gamma=\mathcal{Z}=\{S_i\}'\cap\{S_i\}''=M_\gamma\cap M_\gamma'$ and is generated by the
two orthogonal projections
\begin{equation}
\label{singlet}
P=\frac{1}{4}\Big(\mathbbm{1}-T\Big)\ ,\quad Q=\mathbbm{1}-P=\frac{1}{4}\Big(3+T\Big)\ ,
\end{equation}
where the first one is $1$-dimensional and  projects onto the two-qubit singlet state
\begin{equation}
\label{singlet1}
\vert \Psi\rangle=\frac{1}{\sqrt{2}}\Big(\vert 0\rangle\otimes\vert 1\rangle\,-\,
\vert 1\rangle\otimes\vert 0\rangle\Big)\ ,
\end{equation}
with $\sigma_3\vert 0\rangle=\vert 0\rangle$ and $\sigma_3\vert 1\rangle=-\vert 1\rangle$.

From Remark $2.$, $\rho_\infty^{\otimes 2}=\rho^*_\infty\otimes\rho^*_\infty$ is a
full-rank stationary state; then,~(\ref{limitstate}) ensures that the asymptotic state
$\rho_\infty$ corresponding to an initial $\rho$ is obtained as $\mathbb{E}[\rho]$,
by means of~(\ref{dualcond}).
In order to construct it, we first construct the conditional expectation
$\hat{\mathbb{E}}$ onto the sub-algebra of
constant matrices; $\hat{\mathbb{E}}$ must be such that
$\displaystyle \hat{\mathbb{E}}[x]=\lambda(x)\,P+\mu(x)\,Q$.
From the properties~(\ref{expcondprop}) of the conditional expectation,
$$
\hat{\mathbb{E}}[PxP]=\lambda(x)\,P\ ,\quad \hat{\mathbb{E}}[QxQ]=\mu(x)\,Q\ ,
$$
where, with $\rho(x):={\rm Tr}(\rho\,x)$,
$$
\lambda(x)=\frac{{\rm Tr}(P\,\rho^{\otimes 2}_\infty\,P\,x)}{\rho^{\otimes 2}_\infty(P)}\ ,\quad
\mu(x)=\frac{{\rm Tr}(Q\,\rho^{\otimes 2}_\infty\,Q\,x)}{\rho^{\otimes 2}_\infty(Q)}\ .
$$
Then, from~(\ref{singlet1}) one gets
\begin{equation}
\label{singlet2}
\rho^{\otimes 2}_\infty\,P=\frac{1-r_\infty^2}{4}\,P\ ,\quad r_\infty=\frac{b}{a}\ ,
\end{equation}
so that, given any initial state $\rho$, its asymptotic state $\rho_\infty$ is given by
(compare with~\cite{BFrev})
\begin{eqnarray}
\nonumber
\rho_\infty=\mathbb{E}[\rho]&=&\frac{4\,\rho(P)}{1-r_\infty^2}\,P\rho_\infty^{\otimes 2}\,P\,+\,
\frac{4\,\rho(Q)}{3+r_\infty^2}\,Q\rho_\infty^{\otimes 2}\,Q\\
\label{2qubitexp}
&=&\frac{4(1-\rho(P))}{3+r^2_\infty}\,\rho^{\otimes 2}_\infty\,+\,
\frac{4\rho(P)-1+r^2_\infty}{3+r^2_\infty}\,P\ .
\end{eqnarray}

The entanglement content of any two-qubit state $\rho$ is quantified by the concurrence $C(\rho)$~\cite{Wootters}:
consider the complex conjugate matrix $\rho^*$, construct $\widetilde{\rho}=\sigma_2\otimes\sigma_2\,\rho^*\,\sigma_2\otimes\sigma_2$ and compute the (positive) eigenvalues
$\lambda^2_i$ of $\rho\widetilde{\rho}$. Then,
$C(\rho)=\max\{0,\lambda_1-\lambda_2-\lambda_3-\lambda_4\}$.
For all asymptotic states $\rho_\infty$ in~(\ref{2qubitexp}), one easily calculates
\begin{equation}
\label{2qubitconc}
C(\rho_\infty)=\frac{1}{2(3+r^2_\infty)}\,
\max\Big\{0,2\left|4\rho(P)-(1-r^2_\infty)\right|-2(1-\rho(P))(1-r^2_\infty)\Big\}
\end{equation}

In~\cite{BFrev} the entanglement capability of the environment has been studied by comparing the concurrence
of certain initial states with that of their asymptotes; in the following, we shall focus upon the following
one-parameter family of initial conditions
\begin{equation}
\label{in-stat}
\rho(\alpha)=\alpha\,\mathbbm{1}\,+\,(1-4\,\alpha)\,P\ ,\quad 0\leq\alpha\leq 1/3\ .
\end{equation}
One easily finds that $C(\rho(\alpha))=\max\{0,1-6\alpha\}$.
Furthermore, if
\begin{equation}
\label{asympt-conc0}
0\leq\alpha<\alpha(r_\infty)=\frac{3+r^2_\infty}{6(3-r^2_\infty)}\ ,
\end{equation}
where $\alpha(r_\infty)$ is an increasing
function of $r_\infty$: $1/6\leq\alpha(r_\infty)\leq 1/3$,
the corresponding asymptotic states obtained, according to~(\ref{limitstate}), as
\begin{equation}
\label{asympt-st}
\rho_\infty(\alpha)=\mathbb{E}[\rho(\alpha)]=\frac{12\alpha}{3+r^2_\infty}\,\rho^{\otimes 2}_\infty\,+\,\frac{3+r^2_\infty-12\alpha}{3+r^2_\infty}\, P
\end{equation}
have concurrence
\begin{equation}
C(\rho_\infty(\alpha))=\frac{1}{2}-3\alpha\,\frac{3-r^2_\infty}{3+r^2_\infty}\,>\,0\ .
\label{asympt-conc}
\end{equation}
Otherwise, namely for $\alpha(r_\infty)\leq\alpha$, $\rho_\infty(\alpha)$ is separable.
One can then conclude:
\begin{enumerate}
\item
both $\rho(\alpha)$ and $\rho_\infty(\alpha)$ are separable if
\begin{equation}
\label{cond1}
\frac{1}{6}\,\leq\,\alpha(r_\infty)\,\leq\,\alpha\,\leq\,\frac{1}{3}\ .
\end{equation}
\item
$\rho(\alpha)$ is separable and $\rho_\infty(\alpha)$ is entangled if
\begin{equation}
\label{cond1a}
\frac{1}{6}\leq\alpha\leq\,\alpha(r_\infty)\ .
\end{equation}
\item
Since $\alpha(r_\infty)\geq1/6$, it follows that, when $0\leq\alpha<1/6$, the initial state $\rho(\alpha)$
is entangled as well as $\rho_\infty(\alpha)$; the entanglement difference
\begin{equation}
\label{deltaconc0}
\Delta(\alpha):=
C(\rho_\infty(\alpha))-C(\rho(\alpha))=9\,\alpha\,\frac{1+r^2_\infty}{3+r^2_\infty}\,-\,\frac{1}{2}
\end{equation}
becomes positive (entanglement gain) if
\begin{equation}
\label{deltaconc}
\alpha>\alpha^*(r_\infty)=\frac{3+r^2_\infty}{18(1+r^2_\infty)}\ ,
\end{equation}
where $\alpha^*(r_\infty)$ is a monotonically decreasing function of $r_\infty$: $1/6\geq\alpha^*(r_\infty)\geq1/9$.
\end{enumerate}

\subsection{Three qubits}

As for one and two qubits, in order to fully characterize the set of asymptotic states,
one needs the conditional expectation~(\ref{condexp}); differently from two qubits, in the case of three qubits
its complete expression is still escaping us.
Indeed, the commutant $M_\gamma$ is not commutative and cannot coincide with its center,
$M_\gamma\neq \mathcal{Z}$ (see Appendix A);
neither does $M_\gamma$ coincide with the commutant of its center ($M_\gamma\neq\mathcal{Z}'$), which is the other case where one would immediately know how to construct the conditional expectation~\cite{Fri1,Fri2}.
What can be analytically constructed is at least the action of $\mathbb{E}$ on certain subsets of initial states.

In appendix A, it is showed that the commutant set $\{S_i\}'=M_\gamma$ is the linear span of the
the $3\times 3$ identity matrix and of the following operators
\begin{equation}
\label{tools1}
S^{(ab)}=\sum_{i=1}^3\sigma^{(a)}_i\sigma^{(b)}_i\ ,\quad a<b=2,3\ ;\quad
S=\sum_{i,j,k=1}^3\varepsilon_{ijk}\,\sigma^{(1)}_i\sigma^{(2)}_j\sigma^{(3)}_k\ .
\end{equation}
Further,
the center $\mathcal{Z}$ surely contains the operator
\begin{equation}
\label{opT}
T=\sum_{a<b=2}^3S^{ab}=S^{(12)}+S^{(23)}+S^{(13)}\ .
\end{equation}
Also, the operators
\begin{equation}
\label{A7a}
P^{(ab)}=\frac{\mathbbm{1}-S^{(ab)}}{4}\in M_\gamma
\end{equation}
are projections such that $[P^{(ab)}\,,\,S_i]=0$ for all $i=1,2,3$.
Using~(\ref{KrausME}), given any $\rho\in\iS(M)$,
the states
\begin{equation}
\label{stateab}
\rho^{(ab)}=\frac{P^{(ab)}\,\rho\,P^{(ab)}}{\rho(P^{(ab)})}
\end{equation}
are such that
\begin{equation}
\label{tool2}
L[\rho^{(ab)}]=P^{(ab)}\,L[\rho]\,P^{(ab)}\Longrightarrow \gamma_t[\rho^{(ab)}]= P^{(ab)}\,
\gamma_t[\rho]\,P^{(ab)}\ .
\end{equation}
Moreover, as $P^{(ab)}=|\Psi_{ab}\rangle\langle\Psi_{ab}|\,\mathbbm{1}_c$ projects onto the singlet vector state
$|\Psi_{ab}\rangle$ of the qubits $a$ and $b$, then
\begin{equation}
\label{tool3}
\gamma_t[\rho_{ab}]= P^{(ab)}\rho^{(c)}_t\ ,
\end{equation}
where $\rho^{(c)}_t$ is a state of the qubit $c$.
\medskip

\noindent
\textbf{Proposition 1}\quad
The state $\rho^{(c)}_t$ evolves in time according to the master equation~(\ref{ME2}) for one qubit and
$\mathbb{E}[\rho^{(ab)}]=P^{(ab)}\rho^{(c)}_\infty$, where $\rho^{(c)}_\infty=\rho^*_\infty$ in~(\ref{blochstat}).
\medskip

\noindent
\textbf{Proof:}\quad
The time-evolution of $\rho^{(c)}_t$ is obtained by tracing over the qubits $a$ and $b$ the
expression~(\ref{tool3}) multiplied by $P^{(ab)}$; by using~(\ref{tool2}) one gets:
$$
\partial_t\rho^{(c)}_t={\rm Tr}_{ab}\Big(P^{(ab)}\,L[P^{(ab)}\,\rho^{(c)}_t]\Big)\ .
$$
By splitting the generator as $L=\sum_{p,q=1}^3L_{pq}$, one gets
\begin{eqnarray*}
L[P^{(ab)}\,\rho^{(c)}_t]&=&\Big(L_{aa}+L_{bb}+L_{ab}+L_{ba}\Big)
\Big[|\Psi_{ab}\rangle\langle\Psi_{ab}|\Big]\,\rho^{(c)}_t
\\
&+&
\Big(\underbrace{L_{ac}+L_{ca}+L_{cb}+L_{bc}}_{L_{II}}\Big)\Big[P^{(ab)}\rho^{(c)}_t\Big]\,\rho^{(c)}_t\\
&+&
P^{(ab)}\,L_{cc}[\rho^{(c)}_t]\ .
\end{eqnarray*}
The first contribution vanishes for it consists of the generator of the master equation~(\ref{ME2})
for two qubits acting on the projection onto the singlet state; from~(\ref{2qubitexp}),
this state is stationary and the statement follows.

Since $P^{(ab)}\in\{S_i\}'$, the trace over the qubits $a$ and $b$ of the second contribution
multiplied by $P^{(ab)}$ reads
$$
{\rm Tr}\Big(P^{(ab)}\,L_{II}[P^{(ab)}\,\rho^{(c)}_t]\Big)=
{\rm Tr}\Big(P^{(ab)}\,L_{II}[\mathbbm{1}^{(ab)}\,\rho^{(c)}_t]\Big)\ .
$$
This piece vanishes, too; indeed, all Kraus operators contribute with
terms of the form ${\rm Tr}(P^{(ab)}\,\sigma_i^{(a)})$ or ${\rm Tr}(P^{(ab)}\,\sigma_i^{(b)})$ which are both zero
as the partial trace of $P^{(ab)}$ is proportional to the $2\times 2$ identity matrix and the Pauli matrices are traceless.
Therefore, $\partial_t\rho^{(c)}_t=L_{cc}[\rho^{(c)}_t]$ whence the result follows from the fact that
$L_{cc}$ is the generator in~(\ref{ME2}) for a single qubit which has $\rho_\infty^*$ as full rank stationary
state.
\medskip

One can now fix the action on the projectors $\displaystyle P^{(ab)}=\frac{\mathbbm{1}-S^{(ab)}}{4}$ and
\begin{equation}
\label{A7b}
P=\frac{2}{3}\sum_{a<b=2}^3P^{(ab)}
\end{equation}
of the dual map $\mathbb{E}$ introduced in~(\ref{dualcond}) which, according to~(\ref{limitstate}), associates to any initial condition the asymptotic states towards which it tends when $t\to+\infty$.
\medskip

\noindent
\textbf{Corollary 1}\quad
$\displaystyle \mathbb{E}[P^{(ab)}]=2\,P^{(ab)}\rho^{(c)}_\infty$ and
$\displaystyle \mathbb{E}[P]=\frac{4}{3}\sum_{a<b=2}^3P^{(ab)}\rho^{(c)}_\infty$.
\medskip

\noindent
\textbf{Proof:}\quad
Set $\rho=P^{(ab)}$ in~(\ref{tool3}); then,
$\displaystyle
\mathbb{E}[P^{(ab)}]=\lim_{t\to+\infty}\gamma_t[P^{(ab)}]=2\,P^{(ab)}\rho^{(c)}_\infty$.
The second relation follows by the linearity of $\mathbb{E}$.
\medskip

\noindent
\textbf{Remark 3.}\quad
Notice that while $P^{(ab)}\in M_\gamma$ and thus $\hat{\mathbb{E}}[P^{(ab)}]=P^{(ab)}$,
$\displaystyle\frac{P^{(ab)}}{2}$ is not an invariant state:
$\displaystyle \mathbb{E}\left[\frac{P^{(ab)}}{2}\right]\neq \frac{P^{(ab)}}{2}$.
\medskip

The last necessary tool for the applications to be discussed in the next section is the action
of the map $\mathbb{E}$ on the projection $Q=\mathbbm{1}-P\in M_\gamma$.
\medskip

\noindent
\textbf{Proposition 2}\quad
$\displaystyle \mathbb{E}[Q]=\frac{8}{1+r^2_\infty}\Big(\rho^{\otimes 3}-\frac{1-r^2_\infty}{6}\sum_{a<b=2}^3
P^{(ab)}\rho^{(c)}_\infty\Big)$.
\medskip

\noindent
\textbf{Proof:}\quad
Since $Q\in M_\gamma$, the properties~(\ref{expcondprop}) and the algebraic relations~(\ref{A8}) applied to
$\hat{\mathbb{E}}[x]=\lambda(x)\,\mathbbm{1}+\sum_{a<b=2}^3\lambda_{ab}(x)\,S^{(ab)}+\mu(x)\,S$,
$x\in M$, give
$$
\hat{\mathbb{E}}[Q\,x\,Q]=Q\,\hat{\mathbb{E}}[x]\,Q=\beta(x)\,Q\ ,\qquad\beta(x)=\lambda(x)+\sum_{a,b=1}^3\lambda_{ab}(x)\ .
$$
Using the time-invariant state $\rho^{\otimes 3}_\infty$, $\mathbb{E}[\rho^{\otimes 3}_\infty]=\rho^{\otimes 3}_\infty$,
one obtains
\begin{eqnarray*}
{\rm Tr}\Big(Q\,\rho^{\otimes 3}_\infty\,Q\,x\Big)&=&{\rm Tr}\Big(\rho^{\otimes 3}_\infty\,Q\,x\,Q\Big)=
{\rm Tr}\Big(\mathbb{E}[\rho^{\otimes 3}_\infty]\,Q\,x\,Q\Big)=
{\rm Tr}\Big(\rho^{\otimes 3}_\infty\,\hat{\mathbb{E}}[Q\,x\,Q]\Big)=
\\
&=&{\rm Tr}\Big(\rho^{\otimes 3}_\infty\,\hat{\mathbb{E}}[Q\,x\,Q]\Big)=
\beta(x)\,{\rm Tr}(\rho^{\otimes 3}_\infty\,Q)\ .
\end{eqnarray*}
This gives $\displaystyle \beta(x)=\frac{{\rm Tr}\Big(Q\,\rho^{\otimes 3}_\infty\,Q\,x\Big)}
{{\rm Tr}(\rho^{\otimes 3}_\infty\,Q)}$; on the other hand, for all $x\in M$,
$$
{\rm Tr}\Big(x\,\mathbb{E}[Q]\Big)={\rm Tr}\Big(\hat{\mathbb{E}}[Q\,x\,Q]\Big)=\frac{{\rm Tr}(Q)}{{\rm Tr}(\rho^{\otimes 3}_\infty\,Q)}\,{\rm Tr}\Big(x\,Q\,\rho^{\otimes 3}_\infty\,Q\Big)\ .
$$
Then, the result follows using that (see~(\ref{singlet2}))
$$
P\rho^{\otimes 3}_\infty=\frac{2}{3}\sum_{a<b=2}^3P^{(ab)}\rho^*_\infty\otimes\rho^*_\infty
\otimes\rho^*_\infty=
\frac{1-r^2_\infty}{6}\sum_{a<b=2}^3P^{(ab)}\rho^{(c)}_\infty=\rho^{\otimes 3}_\infty\,P\ .
$$

In the next section, we study the following protocol:
\begin{itemize}
\item
add a third completely depolarized qubit to a two qubit initial state $\rho(\alpha)$ as in~(\ref{in-stat});
\item
let the resulting three qubit state reach equilibirum under the time-evolution governed by~(\ref{ME2});
\item
eliminate from the asymptotic state the added third qubit.
\end{itemize}
We show that the resulting two qubit state
\begin{enumerate}
\item
can be entangled when the asymptotic state reached by the two qubits evolving alone would not,
that is when $\alpha\geq\alpha(r_\infty)$;
\item
can be more entangled than the initial state, when the asymptotic state of  the two qubits evolving alone would not, namely when
$\alpha\leq \alpha^*(r_\infty)$;
\item
the entanglement gain can be larger than $\Delta(\alpha)>0$ when $\alpha>\alpha^*(r_\infty)$.
\end{enumerate}

\section{Applications}

We now apply the previous results to the study of the asymptotic entanglement properties of a class
of three-qubit states obtained by appending to the two-qubit states~(\ref{in-stat}) a third qubit in the completely
depolarized state; we shall thus focus onto initial density matrices of the form
\begin{equation}
\label{thestates}
\rho^{123}(\alpha)=\rho(\alpha)\otimes\frac{\mathbbm{1}}{2}=
\frac{\alpha}{2}\,\mathbbm{1}\,+\,\frac{1-4\,\alpha}{2}\,P^{(12)}\ ,\quad 0\leq\alpha\leq 1/3\ ,
\end{equation}
where, according to the notation of the previous section, $P\otimes\mathbbm{1}=P^{(12)}$.

The corresponding asymptotic states are given by the map $\mathbb{E}:\iS(M)\mapsto\iS(M)$ whose action is given
by Corollary 1 and Proposition 2; indeed, writing $\mathbbm{1}=P+Q$,
\begin{eqnarray}
\label{thestates1}
\mathbb{E}[\mathbbm{1}]&=&\frac{8}{1+r^2_\infty}\,\rho^{\otimes 3}_\infty\,+\,\frac{8r^2_\infty}{3(1+r^2_\infty)}\,
\sum_{a<b=2}^3P^{(ab)}\rho^{(c)}_\infty\\
\nonumber
\rho^{123}_\infty(\alpha)=\mathbb{E}[\rho^{123}(\alpha)]&=&\frac{4\,\alpha}{1+r^2_\infty}\,\rho^{\otimes 3}_\infty\,
+\,\frac{4\,\alpha\,r^2_\infty}{3(1+r^2_\infty)}\,\sum_{a<b=2}^3P^{(ab)}\rho^{(c)}_\infty\\
\label{thestates2}\,
&+&
(1-4\,\alpha)\,P^{(12)}\rho^{(3)}_\infty\ .
\end{eqnarray}

According to the last step of the protocol described at the end of the previous section,
we trace the asymptotic states $\rho^{123}_\infty(\alpha)$ with respect to the appended qubit:
\begin{eqnarray}
\nonumber
\rho^{12}_\infty(\alpha)&=&{\rm Tr}_3(\rho^{123}_\infty(\alpha))=
\frac{4\,\alpha}{1+r^2_\infty}\,\rho_\infty\otimes\rho_\infty\,+\,
\frac{4\,\alpha\,r^2_\infty+3(1-4\,\alpha)(1+r^2_\infty)}{3(1+r^2_\infty)}\,P\\
\label{thestates3}
&+&\frac{2\,\alpha\,r^2_\infty}{3(1+r^2_\infty)}\,\Big(\mathbbm{1}\otimes\rho_\infty\,+\,
\rho_\infty\otimes\mathbbm{1}\Big)\ ,
\end{eqnarray}
where $P$ projects onto the two-qubit singlet state.
The concurrence $C(\rho^{(12)}_\infty(\alpha))$ of this two-qubit state can be computed and compared with that of
the asymptotic state $\rho^\infty(\alpha)$ in~(\ref{deltaconc}); though easy to calculate, the expression of the concurrence is not particularly inspiring and can be found in Appendix B, equation~(\ref{B3}).
The goal is to see whether
the addition and final discarding of the added qubit may increase the asymptotic entanglement of
$\rho_\infty(\alpha)$ in~(\ref{asympt-st}).

We start by considering the case of separable two-qubit state $\rho(\alpha)$ that cannot get asymptotically
entangled by the action of the master equation~(\ref{ME2}).
According to~(\ref{cond1}), this occurs for
$1/6\,\leq\,\alpha(r_\infty)\,\leq\,\alpha\,\leq\,1/3$.

Consider a third qubit prepared in the totally depolarized state and appended to the qubits $1$ and $2$ prepared in a
state $\rho(\alpha)$ with $\alpha$ in the above range.
According to Appendix B, by tracing the asymptotic $3$-qubit state $\rho^{123}_\infty(\alpha)$ over
the appended qubit, the qubits $1$ and $2$ are entangled, that is their concurrence $C(\rho^{12}(\alpha))>0$,
if either $0\leq\alpha\leq \alpha_+(r_\infty)$ where
$$
\alpha_+(r_\infty)\,=\,\frac{3(1+r^2_\infty)}{4(3+2\,r^2_\infty)
+2\sqrt{\delta(r_\infty)}}\ ,
$$
or $\displaystyle\alpha_-(r_\infty)\leq\alpha\leq\frac{1}{3}$ where
$$
\alpha_-(r_\infty)\,=\,\frac{3(1+r^2_\infty)}{4(3+2\,r^2_\infty)
-2\sqrt{\delta(r_\infty)}}\ ,
$$
with $\delta(r_\infty)=(1-r^2_\infty)\Big((3+2r^2_\infty)^2-9\,r^2_\infty\Big)\geq 0$.
One checks that $\alpha_+(r_\infty)\leq \alpha(r_\infty)$; therefore, the first condition
is incompatible with~(\ref{cond1}).

We shall then set $r_\infty$ so that $\displaystyle\alpha_-(r_\infty)\leq\alpha\leq\frac{1}{3}$ and
let $0.980965=r^*\leq r_\infty\leq 1$ as calculated in Appendix B, equation~(\ref{B7}).
If $\alpha_-(r_\infty)\leq \alpha(r_\infty)$,
all initial states $\rho(\alpha)$ with $\alpha>\alpha(r_\infty)$ correspond to asymptotic states
$\rho_\infty(\alpha)$ which are separable, but to reduced asymptotic states $\rho^{12}_\infty(\alpha)$
in~(\ref{thestates3}) that are entangled.
The same occurs for $\alpha_-(r_\infty)\geq \alpha(r_\infty)$ for initial states
$\rho(\alpha)$ with $\alpha>\alpha_-(r_\infty)$.
Therefore, there are separable states $\rho(\alpha)$ which do not get asymptotically entangled by direct
immersion in the environment described by~(\ref{ME2}), but do get entangled if a third depolarized qubit
is appended to them and then eliminated after reaching stationarity.

This phenomenon is numerically confirmed in the following figure where the concurrence
of $\rho^{12}_\infty(\alpha)$ is plotted for  $0\leq r_\infty\leq 1$ and
$1/5\leq\alpha(r_\infty)\,\leq\,\alpha\,\leq\,1/3$.
\begin{figure}[h!]
\centering
\includegraphics[width=6cm]{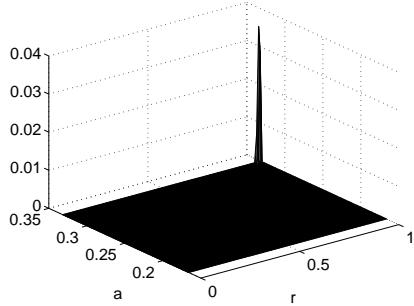}
\caption{$C(\rho^{12}_\infty(\alpha))$ when $C(\rho_\infty(\alpha))=C(\rho(\alpha))=0$, $r=r_\infty$,
$a=\alpha$}
\label{fig1}
\end{figure}
\medskip

\noindent
{\bf Remark 4.}\quad
Admittedly, the range of favorable values of the environment dependent parameter $r_\infty$ is not so large,
as well as the range of separable two-qubit states $\rho(\alpha)$ that can get entangled by means of the protocol
and not by direct immersion in the environment.
However, the fact that such a possibility exists is an indication of what might be achievable if one could
completely characterize the whole manifold of stationary three-qubit states.
Also, instead of tracing away the third qubit, one could perform a less mixing operation on it in such a way
that some more entanglement be localized on the remaining two qubits: preliminary results confirm this
possibility, but, unfortunately, not to a sufficiently significative extent.
\medskip

Luckily, concerning the second two points listed at the end of Section 3.3, addition of a third
completely depolarized qubit and its elimination after reaching the stationary regime, allows for a more
substantial improvement on the entanglement that can be gained asymptotically.
Let us consider the difference
\begin{equation}
\label{D1}
\Delta_1(\alpha):=C(\rho_\infty^{12}(\alpha))\,-\,C(\rho(\alpha))
\end{equation}
in the range $\alpha<\alpha^*(r_\infty)$.
For these values of $\alpha$, no entanglement gain can be achieved by letting the two open qubits evolve
towards their stationary state; that is,
$$
\Delta(\alpha)=C(\rho_\infty(\alpha))-C(\rho(\alpha))\leq0\ .
$$
However, by adding a completely depolarized qubit and eliminating it after reaching
the stationary state, one may get $\Delta_1(\alpha)>0$ as showed in Figure 2 which exhibits the range of parameters $r_\infty$ (depending on the environment) and $\alpha$ (labeling the initial state) for which this is possible.
\begin{figure}[h!]
\centering
\includegraphics[width=6cm]{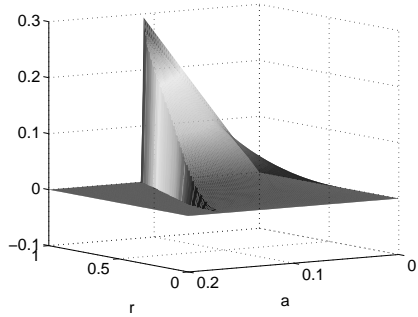}
\caption{$\Delta_1(\alpha)$ vs $r=r_\infty$, $0\leq a=\alpha\leq \alpha^*(r_\infty)\leq 1/6$}
\label{fig2}
\end{figure}

Next, consider the difference
\begin{equation}
\label{D2}
\Delta_2(\alpha):=C(\rho_\infty^{12}(\alpha))\,-\,C(\rho_\infty(\alpha))
\end{equation}
in the range $\alpha>\alpha^*(r_\infty)$ where two qubits present an entanglement gain, $\Delta(\alpha)>0$.
Such an entanglement gain may be increased by adding a third depolarized qubit as
showed in Figure 3.
\begin{figure}[h!]
\centering
\includegraphics[width=6cm]{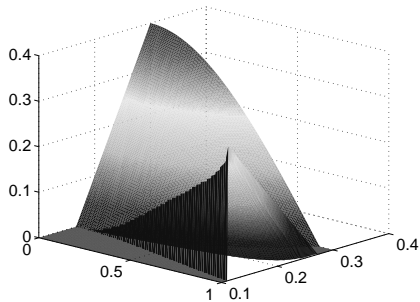}
\caption{$\Delta_2(\alpha)$ vs $r=r_\infty$, $\alpha^*(r_\infty)\leq a=\alpha\leq 1/3$}
\label{fig3}
\end{figure}

\section{Conclusions}

We have studied the asymptotic states of a Lindblad master equation describing the reduced dynamics
of three qubits weakly coupled to an environment that affects in the same way any pair of qubits.
By applying standard algebraic techniques, we could control the asymptotic states of a particular family
of initial three qubit states of which one is completely depolarized.
We showed that, after eliminating the latter from the asymptotic state, the remaining two qubits may show
more entanglement than the asymptotic two-qubit state achievable by direct immersion within
such an environment.
This phenomenon can be regarded as an asymptotic manifestation of the richer structure of irreversible entanglement generation in higher dimensional discrete systems that was observed at short times in~\cite{BLN}.

\section*{Appendix A}

Given the operators $S_i=\sum_{a=1}^3\sigma^{(a)}_i$, $i=1,2,3$, the commutant set $M_\gamma=\{S_i\}'$ is found by expanding a generic $x\in M$ by means of tensor products of the Pauli matrices:
\begin{equation}
\label{A1}
x=\lambda_0\mathbbm{1}+\sum_{a=1,i=1}^3\lambda^{(1)}_{ai}\sigma^{(a)}_i+
\sum_{a<b=2;i,j=1}^3\lambda^{(2)}_{ai,bj}\sigma^{(a)}_i\sigma^{(b)}_j+
\sum_{i,j,k=1}^3\lambda^{(3)}_{ijk}\sigma^{(1)}_i\sigma^{(2)}_j\sigma^{(3)}_k\ ,
\end{equation}
and then imposing $[x\,,\,S_i]=0$ for all $i=1,2,3$. By using the Pauli algebraic relations
one finds the following equalities
\begin{eqnarray}
\label{A2a}
&&
\sum_{\ell=1}^3\lambda^{(1)}_{a\ell}\,\varepsilon_{\ell pi}=0\qquad\forall\ i,p=1,2,3\\
\label{A2b}
&&
\sum_{\ell=1}^3\Big(
\lambda^{(2)}_{ai,b\ell}\,\varepsilon_{\ell pj}+\lambda^{(2)}_{a\ell,bj}\varepsilon_{\ell pi}\Big)=0
\qquad\forall\ a<b=2,3\,;\,i,j,p=1,2,3\\
\label{A2c}
&&
\sum_{\ell=1}^3\Big(\lambda^{(3)}_{ij\ell}\,\varepsilon_{\ell pk}+
\lambda^{(3)}_{i\ell k}\,\varepsilon_{\ell pj}+
\lambda^{(3)}_{\ell jk}\,\varepsilon_{\ell pi}\Big)=0\qquad\forall\ i,j,k,p=1,2,3\ ,
\end{eqnarray}
whence $\lambda^{(1)}_{ai}=0$ for all $a,i=1,2,3$, $\lambda^{(2)}_{ai,bi}=\lambda^{(2)}_{aj,bj}$ for all
$a<b=2,3$ and $i,j=1,2,3$, while $\lambda^{(3)}_{ijk}=\lambda\,\varepsilon_{ijk}$.
It thus follows that the commutant set is $\{S_i\}'=\{\mathbbm{1},S^{(ab)},S\}$, $a,b=1,2,3$, namely the linear span of $\mathbbm{1}$ and
\begin{equation}
\label{A3}
S^{(ab)}=\sum_{i=1}^3\sigma^{(a)}_i\,\sigma^{(b)}_i\quad a<b=2,3\ ;\qquad
S=\sum_{i,j,k=1}^3\varepsilon_{ijk}\,\sigma^{(1)}_i\,\sigma^{(2)}_j\,\sigma^{(3)}_k\ .
\end{equation}
Unlike for two qubits, the commutant set is not commutative; indeed, with $a,b,c$ different indices,
\begin{eqnarray}
\label{A4a}
&&
\Big[S^{(ab)}\,,\,S^{(ac)}\Big]=2i\varepsilon_{abc}\,S^{(bc)}\ ,\quad
\Big\{S^{(ab)}\,,\,S^{(ac)}\Big\}=2\,S^{(bc)}\\
\label{A4c}
&&
\Big[S^{(ab)}\,,\,S\Big]=4i\Big(S^{(bc)}-S^{(ac)}\Big)\ ,\quad a<b\ ,
\end{eqnarray}
whence
\begin{equation}
\label{A5}
T=\sum_{a<b=2}^3S^{ab}=S^{(12)}+S^{(23)}+S^{(13)}\Rightarrow [T\,,\,S^{(ab)}]=[T\,,\,S]=0\ ,
\end{equation}
so that $T$ belongs to the center $\mathcal{Z}=\{S_i\}'\cap\{S_i\}''=M_\gamma\cap M_\gamma'$.
Other useful algebraic relations are as follows
\begin{equation}
\label{A6}
(S^{(ab)})^2=3-2\,S^{(ab)}\ ,\quad a,b=1,2,3\ ;\qquad  S^2=2(3-T)\ .
\end{equation}
From the first relations it follows that
\begin{eqnarray}
\label{A7aa}
P^{(ab)}&=&\frac{\mathbbm{1}-S^{(ab)}}{4}\in M_\gamma=\{S_i\}'=\{\mathbbm{1},S^{(ab)},S\}\\
\label{A7bb}
P&=&\frac{2}{3}\sum_{a<b=2}^3P^{(ab)}=\frac{1}{2}\Big(\mathbbm{1}-\frac{1}{3}\,T\Big)\in M_\gamma\cap M_\gamma'
\end{eqnarray}
are two-dimensional, respectively four-dimensional projections.
In particular, $P^{(ab)}$ is the tensor product of the projection onto the singlet state of the qubits
$a$ and $b$ with the identity matrix for the qubit $c$.
Furthermore, the projection $Q=\mathbbm{1}-P\in M_\gamma\cap M_\gamma'$ fulfils
\begin{equation}
\label{A8}
Q\,S^{(ab)}\,=\,Q\quad\forall\, a<b\ ;\qquad Q\,S=0\ .
\end{equation}
Other projections commuting with $M_\gamma$, that is in the commutant $M_\gamma'$ are thus all
sub-projections $q\leq Q$ for which $qQ=Q=Qq$, whence
$$
q\,S^{(ab)}=q\,Q\,S^{(ab)}=q\,Q=q\ ,\qquad q\,S=q\,Q\,S=0\ .
$$
However, unless $q=Q$, these projections $q$ cannot belong also to $M_\gamma$; this is proved
by writing $q=\lambda\mathbbm{1}+\sum_{a<b=2}^3\lambda_{ab}\,S^{(ab)}+\mu\,S$ and by imposing the previous
conditions.
It thus follows that, for the three-qubit case discussed in this work,
neither $\mathcal{Z}=M_\gamma$ as this would imply $M_\gamma$ commutative, or
$\mathcal{Z}'=M_\gamma$ as this would imply $M_\gamma'\subseteq M_\gamma$.
These are the two conditions for which a conditional expectation onto $M_\gamma$ could easily be
explicitly written~\cite{Fri1,Fri2}.

\section*{Appendix B}

The explicit form of the states $\rho^{12}_\infty(\alpha)$ in~(\ref{thestates3}) is
\begin{equation}
\label{B1}
\rho^{12}_\infty(\alpha)=\frac{1}{1+r^2_\infty}\,
\begin{pmatrix}
x_+&0&0&0\cr
0&y&-u&0\cr
0&-u&y&0\cr
0&0&0&x_-
\end{pmatrix}\ ,
\end{equation}
where
\begin{eqnarray}
\label{B2}
x_\pm&=&\frac{\alpha}{3}\,(1\pm r_\infty)(3(1\pm\,r_\infty)+2r^2_\infty)\ ,\ y=\frac{3(1+r^2_\infty)-2\,\alpha(3+5\,r^2_\infty)}{6}\\
u&=&\frac{3(1+r^2_\infty)-4\alpha\,(3+2\,r^2_\infty)}{6}\ .
\label{B3}
\end{eqnarray}
The concurrence of such a state is
\begin{eqnarray}
\nonumber
&&
C(\rho^{12}_\infty(\alpha))=\frac{1}{3(1+r^2_\infty)}\max\Bigg\{0\,,\,
\Big|
3(1+r^2_\infty)-4\alpha(3+2\,r^2_\infty)
\Big|\\
&&\hskip 6cm
-2\,\alpha\,\sqrt{(1-r^2_\infty)(9+9\,r^2_\infty+4\,r^4_\infty)}\Bigg\}\ .
\label{B4}
\end{eqnarray}
More explicitly, set $\delta(r_\infty)=(1-r^2_\infty)\Big((3+2r^2_\infty)^2-9\,r^2_\infty\Big)\geq0$; then,
\begin{eqnarray}
\label{B4a}
C(\rho^{12}_\infty(\alpha))&=&
\frac{2\,\alpha\,\Big(2(3+2\,r^2_\infty)-\sqrt{\delta(r_\infty)}\Big)}{3(1+r^2_\infty)}-1\\
\nonumber
&\textrm{if}&\quad
\frac{1}{3}\,\geq\,\alpha\,>\,\alpha_-(r_\infty)\,=\,\frac{3(1+r^2_\infty)}{4(3+2\,r^2_\infty)
-2\sqrt{\delta(r_\infty)}}\\
\label{B4b}
C(\rho^{12}_\infty(\alpha))&=&1-
\frac{2\,\alpha\,\Big(2(3+2\,r^2_\infty)+\sqrt{\delta(r_\infty)}\Big)}{3(1+r^2_\infty)}\\
\nonumber
&\textrm{if}&\quad
0\leq\,\alpha\,<\,\alpha_+(r_\infty)\,=\,\frac{3(1+r^2_\infty)}{4(3+2\,r^2_\infty)
+2\sqrt{\delta(r_\infty)}}\ .
\end{eqnarray}
The lower bound $\alpha_-(r_\infty)$ is a decreasing function,
\begin{equation}
\label{B5}
\frac{3}{10}=\alpha_-(1)\leq\alpha_-(r_\infty)\leq\frac{1}{2}=\alpha_-(0)\ ,
\end{equation}
while the upper bound $\alpha_+(r_\infty)$ monotonically increases,
\begin{equation}
\label{B6}
\frac{1}{6}=\alpha_+(0)\leq\alpha_+(r_\infty)\leq\frac{3}{10}=\alpha_+(1)\ .
\end{equation}
While $\alpha_+(r_\infty)$ is always in the permitted range $\displaystyle 0\leq\alpha\leq\frac{1}{3}$,
it turns out that
\begin{equation}
\label{B7}
\alpha_-(r_\infty)\leq \frac{1}{3}\qquad\textrm{if}\qquad 0.980965=r^*\leq r_\infty\leq 1\ ,
\end{equation}
where
$r^*$ is such that $\displaystyle\alpha_-(r^*)=1/3$.

\end{document}